\documentclass[12pt]{article}
\usepackage{amsmath,amssymb,amsfonts}
\usepackage[paper=letterpaper,margin=1.0in]{geometry}
\usepackage{graphicx}

\newcommand{\be}{\begin{equation}}
\newcommand{\ee}{\end{equation}}
\newcommand{\beq}{\begin{eqnarray}}
\newcommand{\eeq}{\end{eqnarray}}
\newcommand{\bea}{\begin{eqnarray}}
\newcommand{\eea}{\end{eqnarray}}
\newcommand{\beqn}{\begin{eqnarray}}
\newcommand{\eeqn}{\end{eqnarray}}

\newcommand{\X}{\mathbb{X}}

\def\pa{\partial}

\newcommand{\rd}{\mathrm{d}}

\renewcommand{\thefootnote}{\fnsymbol{footnote}}
\begin{document}

\begin{center}

\vskip 1.0cm
\centerline{\Large\bf Quantum Gravity, Dynamical Phase Space and String Theory}
\vskip 1.0cm

\centerline{\bf
Laurent Freidel${}^{1}$\footnote{lfreidel@perimeterinstitute.ca},
Robert G. Leigh${}^{2}$\footnote{rgleigh@illinois.edu} and
Djordje Minic${}^{3}$\footnote{dminic@vt.edu} 
}

\vskip 0.5cm
 
{\it
${}^1$Perimeter Institute for Theoretical Physics, 31 Caroline St. N., Waterloo ON, N2L 2Y5, Canada \\
\vskip 0.2cm
${}^2$Department of Physics, University of Illinois, 1110 W. Green St., Urbana IL 61801, USA \\
\vskip 0.2cm
${}^3$Department of Physics, Virginia Tech, Blacksburg VA 24061, USA \\
\vskip 0.2cm
}

\vskip 1.0cm
\centerline{Submission date: March 30, 2014}
\vskip 1.0cm

\begin{abstract}
In a natural extension of the relativity principle we argue that a quantum theory of gravity involves two fundamental scales
associated with both dynamical space-time as well as dynamical momentum space. This view of quantum gravity  is explicitly realized in a new formulation of string theory which involves dynamical phase space and in which space-time is a derived concept.
This formulation naturally unifies symplectic geometry of Hamiltonian dynamics, complex geometry of quantum theory
and real geometry of general relativity. The space-time and momentum space dynamics, and thus dynamical phase space, is governed by a new version of the Renormalization Group.
\end{abstract}

\vspace{1cm}

\centerline{Essay written for the Gravity Research Foundation 2014 Awards for Essays on Gravitation}

\vspace{1cm}

\end{center}

\setcounter{footnote}{0}
\renewcommand{\thefootnote}{\arabic{footnote}}
\newpage

\section{What is Quantum Gravity?}

The problem of quantum gravity is, arguably,  one of the most outstanding open questions in theoretical physics.
In our recent work \cite{flmprl} we have approached this problem from a novel point of view by invoking a natural 
extension of the relativity principle in the quantum context. 
Recall that Einstein's special theory of relativity famously integrated two seemingly
different concepts, the concept of space and the concept of time, into a new concept of space-time, by using the fundamental constant $c$ as a universal conversion factor between space and time. Furthermore Einstein's general relativity made this novel kinematical vision of space-time fully dynamical, thus providing a new view on the gravitational field. 
In this essay we argue that quantum mechanics forces us to consider 
phase space as the natural arena for physics and that gravity provides a way to unify space-time and momentum space through the universal conversion factor $G_{N}$.
In this unified framework  phase space is fully dynamical.
This is the new structure that we present.

Let us start with a point of clarification: when we talk about phase space, we do not mean the phase space of gravitational fields;
we refer exclusively to the phase space of matter probes.
We can think of these probes as particles or as strings.
The nature of the probe will not matter for the first part of the discussion, before we present more specific results in the case of string theory.

\subsection{Born reciprocity}

The first key point is the simple remark that in quantum mechanics the concepts of space-time and momentum space are naturally unified into phase space (of probe degrees of freedom).
In the classical framework this justifies the use of Hamilton's formulation and of phase space as a space equipped with a symplectic structure denoted as $\omega$. At the quantum level we have a Hilbert space  and  unitary transformations acting on it.
 One of the remarkable properties of these unitary transformations is that they can be used  to relate a space-time basis to a momentum basis and interchange spatial and momentum coordinates
 according to 
$ x_a \to p_a$ and $p_a \to - x_a$. 
This duality map, denoted as $I$,  constitutes a {\it complex structure} 
on phase space, that is, a map acting on tangent space, such that $I^{2}=-1$.
Moreover, as a unitary transformation it is also a symplectic transformation preserving $\omega$, i.e., $I^{T}\omega I=\omega$.
This means that $H \equiv \omega I $ is a non-degenerate metric on phase space.
In summary,
one of the striking and inevitable properties of the quantum theory is to provide  phase space with a complex structure and a {\it generalized metric} $H_{AB}$.
The complex structure is intimately related to  the Hilbert space structure, while the metric is intimately related to the probabilistic measure of quantum mechanics \cite{kibble}.
 
One of the central insights of Born, 75 years ago \cite{born}, was to notice that 
the kinematical symmetry generated by $I$ is fundamentally broken 
in the presence of gravity.
Indeed, general relativity explicitly breaks this symmetry because it states that space-time is curved, 
while energy-momentum space, defined as a cotangent space, is linear and flat. 
It is important to appreciate that this kinematical breaking is different from the dynamical breaking that arises in Schr\"odinger's quantum mechanics, due to the choice of a particular Hamiltonian. In the latter case we can still change the basis of states. In the presence of gravity this is no longer possible\footnote{Apart from homogeneous manifolds, there is no general theory of  Fourier transformation in the curved context.}. There is a preferred basis --- the space-time basis.
According to Born, the clash between the kinematical duality symmetry of quantum mechanics and the presence of dynamical space-time curvature is what prevents us from  a theory of quantum gravity.
In other words, a theory of quantum gravity ought to restore the kinematical duality principle.
Since in Einstein's theory of gravity the space-time metric is dynamical,
Born reciprocity would suggest that 
 the generalized phase space metric $H_{AB}$  has to be dynamical and that one should have to endow momentum space with a dynamical metric structure.

\subsection{Born reciprocity reformulated and completed}

Unfortunately, it has been notoriously difficult to implement Born's intuition. Throughout the last 75 years, a few valiant attempts have been made 
to incorporate momentum space curvature as a natural regulator in quantum field theory, without any definite success
(apart from Born and his collaborators, the efforts of Snyder and Golfand \cite{Golfand} are particularly noteworthy). 
As recognized recently \cite{relatloc}, one of the difficulties  resides in  a rather radical aspect of Born's proposal: allowing for a curved momentum space necessarily implies some amount of non-locality in space-time, which clashes with local quantum field theory.
Another issue stems from the fact that no 
convincing generalization of Einstein's equations to the generalized metric $H$, or a principle behind such equations, has ever been proposed.
As we are going to see, the crucial missing ingredient that completes Born's quantum intuition is provided by another fundamental relation  between space-time and momentum space stemming from gravity \cite{flmprl, flmlong1}. 
Note that the notion of a momentum space metric which follows from quantum mechanics also appears in the context of 
quantum field theory, which merges special relativity with quantum mechanics. Indeed, the mass shell condition requires the presence of a metric in momentum space.
It is also interesting to note that in the case of a space-time of constant curvature, such as de Sitter space, where momentum space can be nonlinearly defined, it is clear that both the space-time and the momentum space geometry of particles are affected by the presence of curvature\footnote{The particle phase space changes from being  an orbit of the Poincar\'e group to an orbit of the de Sitter group. These have widely different geometries.}. This example confirms Born's intuition and 
gives us a theoretical ``smoking gun'' evidence, in the context of constant curvature, that gravity affects the geometry of phase space.

More generally, our second key point
follows from the trivial remark that, in 4 dimensions at least, 
$G_{N}$ has dimensions of length over energy (assuming that $c=1$). Moreover, by the equivalence principle, $G_N$ is a universal coupling for all forms of matter and energy.
This means that $G_{N}$ can be considered as a universal conversion factor between the seemingly different concepts of space and momentum space, in the same way that the speed of light is a universal conversion factor between space and time.
The presence of such a conversion factor allows one to blur the line between space and momentum space and provides a unification of space with momentum space.
 This adds another central insight that was missing in Born's original proposal \cite{flmprl}. Even if  we unify space with momentum space into a generalized geometrical object,  we still need to be able 
to split the phase space into space and momentum space ---  in other words,  we need an extra structure that tells us what is $x_a$ and what is $p_a$.
In technical terms, space-time appears as a Lagrangian sub-manifold of the probe phase space, that is, a manifold of maximal dimension on which the symplectic structure vanishes.
Similarly, momentum space is another Lagrangian sub-manifold, and these two sub-manifolds intersect transversely. In other words, there is no element which can be at the same time both space and momentum.
This structure of two transverse Lagrangian sub-manifolds in a phase space  defines  a {\it bi-Lagrangian structure} \cite{flmprl}. It is a necessary structure that we constantly use in physics, albeit rather implicitly.
Quite remarkably this structure is entirely encoded in the choice of another metric 
on phase space, denoted as $\eta_{AB}$.
This metric possesses two key  properties: first it is {\it neutral}, that is of signature $(D,D)$ for a space-time of dimension $D$,
and second, the Lagrangian sub-manifolds are null subspaces with respect to $\eta$.
In order to see this, suppose that we have a bi-Lagrangian structure on phase space. We can introduce an operator, denoted $K$, such that
its value on vectors tangent to the space-time Lagrangian is $+1$, while its value on vectors tangent to the momentum space Lagrangian is $-1$.
This operator is\footnote{We do not discuss the integrability conditions in this essay.} a {\it real structure}, that satisfies $ K^{2}=1$. Moreover,  this operator is anti-compatible with the symplectic structure $ K^{T} \omega K =-\omega$ \cite{flmprl}.
This implies that $\eta \equiv \omega K$ is a neutral metric associated with the bi-Lagrangian.
In the usual case this neutral metric is simply the pairing between vectors $x^{a}$ and covectors $p_{a}$. A remark that will be important later is that in the usual case that we implicitly work with,
the real and complex structure anti-commute: $KI=-IK$.
Note that once we have a generalized metric $H$ and a neutral metric $\eta$ we can define the space-time to be a maximally null subspace of $\eta$ and the space-time metric to be given by the restriction of $H$ to this $\eta$-null subspace.

\subsection{Summary: Quantum gravity and dynamical phase space}

In summary, we have seen that both quantum mechanics via the presence of $\hbar$, and gravity through a universal conversion factor $G_N$,  lead to different phase  space unifications. 
The two crucial remarks we have made are that on one hand  quantum theory introduces naturally  a complex structure $I$ ($I^2=-1$) compatible with $\omega$.
On the other hand, the splitting of phase space between space and momentum space introduces naturally a real structure $K$ ($K^2=1$) anti-compatible with $\omega$.
The complex structure leads to a generalized metric $H=\omega I$ while the real structure leads to a neutral metric $\eta = \omega K$
(the $(D,D)$ metric $\eta$ can be heuristically understood as providing a generalized causal structure on phase space).
In pure quantum mechanics, $H$ is a kinematical structure,
while $\eta$ or the choice of preferred polarization, can be modified 
by unitary dynamics.
Alternatively, in pure gravity $\eta$ is a kinematical structure
(space-time provides a preferred polarization that cannot be modified),  while $H$, or at least its space-time part, becomes dynamical.
According to this view, if we introduce gravity in quantum theory, we have to make $H$ dynamical. This is Born's argument \cite{born}. Moreover if we introduce
 quantum theory in gravity, 
we have to make $\eta$ dynamical. This is our addition to Born's argument \cite{flmprl}.
 Since ``quantizing'' gravity or ``gravitizing'' quantum mechanics should result in the same quantum gravity theory,  both $\eta$ and $H$ should be allowed to be dynamical. 
It is important to note that the kinematical nature of $\eta$ in gravity is tied up with the assumption of locality, i.e., the existence of a single well-defined space-time for all matter probes. 
Therefore a dynamical $\eta$ necessarily leads to a relaxation of locality and a breakdown of effective field theory.

To conclude, if one wants to formulate a full theory of quantum gravity one  needs a formulation which contains (if we set the space-time conversion factor $c=1$)  not only a fundamental length scale $\lambda$ but also a fundamental energy scale $\varepsilon$ from which we can reconstruct $\hbar = \lambda \varepsilon$ and $G_{N} \propto \lambda/\varepsilon$ (in $D=4$)
\cite{flmprl, flmlong1}.
Moreover, the purely kinematical splitting of phase space into space and momentum space has to be made dynamical, and thus both the phase space metric $H_{AB}$ and the neutral metric $\eta_{AB}$ have to be dynamical in the context of quantum gravity.
Finally, we have seen that the geometry of quantum gravity naturally  unifies the symplectic geometry of classical mechanics with the complex structure of quantum mechanics together with the metric or real structure of gravity. 

It is also quite remarkable that  the above intuition regarding
 the crucial importance of dynamical momentum space in quantum gravity appears in the context of three different models involving quantum gravity or a well-defined notion of non-locality.
The first example is quantum gravity in 3 space-time dimensions.
In this case, by coupling matter to 3d quantum gravity and integrating out the gravitational degrees of freedom, 
one discovers that the effective theory for the matter \cite{3dg} is fundamentally non-local: one obtains a field theory whose momentum space is homogeneously curved.
The Born reciprocity principle also appears in the context of the renormalization of non-commutative field theories, in which the
fixed point is fundamentally Born reciprocal and the RG flow mixes 
short distance scales with long distance scales \cite{grosse}.
Similarly, it has been argued that dynamical momentum space associated with Born reciprocity implies dramatic phenomenological consequences in the context of the vacuum energy problem \cite{cc}, which also exhibits the mixing of short and long distance scales.
The most surprising aspect of our work \cite{flmprl,flmlong1,chiral}
 is that string theory provides a precise realization of the geometric structure
unifying symplectic, metric and complex aspects of the geometry of quantum
gravity into one unifying whole, together with a specific dynamical principle.

\section{String Theory and Dynamical Phase Space}

In order to see that, we start from a formulation of string theory that naturally incorporates the two scales $\lambda,\epsilon$ and is therefore 
more general than the Polyakov formulation\footnote{The Polyakov action has the dimension of area and requires only one length scale. In this formulation, the momentum scale appears only if we define an effective quasi-local truncation of the string spectrum, but it is not part of the fundamental definition \cite{veneziano}. }. 
This is realized by the Tseytlin action \cite{chiral}, whose target space unifies space-time and momentum space. 
This formulation necessitates the presence of a generalized metric $H=\omega I$ and a neutral metric $\eta=\omega K$ as predicted in the first part of this essay
\be
\frac{1}{\hbar}S= \frac12\int \rd \sigma\rd \tau \left(\partial_{\tau}{\X}^{A} \partial_{\sigma}\X^{B} \eta_{AB}  -   \partial_{\sigma}\X^{A} \partial_{\sigma}\X^{B} H_{AB}\right).
\ee
Here it  is convenient, as suggested by the double field formalism \cite{double}, 
to introduce  dimensionless  coordinates $\X^{A}\equiv (X^{\mu}/\lambda ,P_{\mu}/\varepsilon )^{T}$  on phase space.
Given a pair $(H,\eta)$ it is natural to consider the operator
$J\equiv \eta^{-1}H$. The consistency of string theory requires 
$J$ to be a chiral structure, that is, a real structure ($J^2=1$) compatible with $\eta$, implying that $J$ is an $O(D,D)$ transformation. 
The presence of this chiral structure is due to the fact that unlike particles, strings have both left- and right-moving modes.
These are defined as the eigenstates of $J$ with value $\pm1$.
The dynamical nature of phase space geometry translates into the fact that left-movers and right-movers can experience different geometries.

This formulation does not require the presence of a bi-Lagrangian structure $K$. However when we look at solutions of string theory, we find this extra structure. Indeed, 
string solutions ultimately do propagate in a sub-manifold which is null with respect to $\eta$. This null subspace is the  effective space-time experienced by the string solution and is defined to be the Lagrangian sub-manifold with $K=+1$. In this sense, different strings
propagate in different space-time backgrounds which are made consistent via string interactions and through a dynamical
momentum space \cite{flmlong1}, thus  giving a sharp example of  relative locality \cite{relatloc}.
(Also, this offers an interesting new perspective on the background independent quantization of string theory.)
When this Lagrangian manifold is assumed to be a fixed kinematical $\eta$-null subspace (which requires $\eta$ to be kinematical) then 
we recover the usual Polyakov formulation where all strings propagate in the same fixed Lagrangian subspace, called space-time.
But we see that this special case happens only when $\eta$ is kinematical, and not dynamical.

\subsection{String theory and Born geometry}

The next remarkable effect comes from T-duality \cite{flmprl}.
In the Polyakov framework, T-duality \cite{joep} is an exact symmetry under the exchange of worldsheet space  $\sigma$ with worldsheet time $\tau$.
In the Tseytlin framework, this worldsheet duality  is no longer a symmetry of the worldsheet formulation. Instead T-duality now appears as a target space duality symmetry $\X \to J(\X)$, implemented by the map $J$. Since $J^{2}$=1 and $J$ preserves $\eta$, both $H$ and $\eta$ are invariant under the duality symmetry generated by $J$.
In order to generalize one of the key properties of T-duality, it is natural to demand that $J$ maps the space-time Lagrangian $K=1$ onto the momentum space Lagrangian $K=-1$. Technically this means that we demand that $J$ and $K$  anticommute as an expression of T-duality.
The same geometry can be characterized in terms of a generalized metric $H$ and two neutral metrics $ \eta$ and $\eta I$.
Since $J$ and $K$ anticommute the combination $I=KJ$ is a complex structure \cite{flmprl, flmlong1}.
Therefore we recover dynamically a tightening of the quantum gravity geometrical structure described above. Any string solution comes equipped with what we call a Born structure \cite{flmprl,flmlong1} $(\omega,I,J,K)$, 
where $J$ and $K$ are real structures and $I$ is a complex structure 
and all three are compatible in the sense that $1 = IJK$ and $I$, $J$ and $K$ anticommute with each other.
Born geometry $(\omega,I,J,K)$ can be viewed as
a natural unification of the symplectic geometry of Hamiltonian dynamics, the complex geometry of quantum theory
and the real geometry of general relativity \cite{gibbons}.

This new viewpoint on string theory may be thought of as a dynamical chiral  phase space formulation of string theory, in which Born reciprocity is properly implemented as a choice of a Lagrangian submanifold of the phase space, and amounts to a generalization of T-duality \cite{flmprl}.
Also, in phase space string theory, the usual dynamical space-time picture appears only as an induced concept (the choice of a Lagrangian submanifold in a dynamical bi-Lagrangian structure) on equal footing with
its complementary and likewise induced dynamical momentum space picture (the choice of a dual Lagrangian submanifold in the same dynamical bi-Lagrangian structure).

\subsection{Two scale Renormalization Group}

In field theory the concept of locality is intricately linked to the understanding of renormalization group (RG) flow.
In phase space string theory we expect the presence of {\it two} fundamental scales, i.e. a length scale and a momentum scale,
to radically alter our
understanding of the RG. This modification is similar to the
 known example of non-commutative field theory
in which the target space RG obeys Born reciprocity \cite{grosse}: we encounter a double RG flow \cite{flmlong1}, in which
one flows from the UV towards lower energy scales, and from the IR to the shorter distance scales in order to end up
at a {\it self-dual} fixed point. 
One of the remarkable features of string theory \cite{joep} is that Einstein's equations for the background space-time metric can be derived as the RG equations on the worldsheet. In particular, the string equations of motion are 
obtained by demanding worldsheet conformal invariance which  brings dynamical constraints on the target metric field.
The doubling of the RG that we expect in target space also happens at the worldsheet level. Since there are two modes (left and right) experiencing possibly different geometries, there are now two different conditions of worldsheet conformal invariance to be imposed. So at the quantum level the string equations of motion are obtained by looking at a fixed point of this double RG flow. 
This gives new background field equations for $\eta_{AB}$ and $H_{AB}$,
for both dynamical space-time and dynamical momentum space, which, in turn, consistently define a dynamical phase space.
At the linearized level, these equations can be written schematically as 
\be
\Box_{H}H_{AB} =0,\quad \Box_{H}\eta_{AB}=0,\quad \Box_{\eta}H_{AB} =0,
\ee
where $\Box_{H}\equiv H^{AB}\pa_{A}\pa_{B}$ is the box operator 
for the $H$ metric.
The first two equations show that both metrics $H$ and $\eta$ are indeed dynamical.
 Since schematically $\Box_{\eta}H = \pa_{X}\pa_{P}H$, the last equation essentially says that the phase space metric can depend only on $X$ 
 or only on $P$. In the first case, we recover the usual graviton mode propagation on this space-time.
 At the non-linear level,  unless we force $\eta$ to be flat,
 we have a coupling between the phase space metric $H$ and the causal metric $\eta$.  

\section{Outlook}

Our approach can be understood as the natural synthesis of two major views on quantum gravity: the first one coming from general relativity, physically rooted in the equivalence principle, in which the emphasis is placed on diffeomorphism invariance (or background independence) and the second one coming from quantum field theory and string theory in which space-time geometry emerges as a consistent background of certain probes described by conformally invariant quantum field theory in two dimensions. In our approach the concept of diffeomorphism invariance is extended to involve a dynamical phase space (of matter degrees of freedom), and likewise the two dimensional formulation of string theory is extended to a phase space formulation which naturally admits a curved background momentum space for the same reasons it admits a curved background space-time. 

Perhaps the most remarkable outcome of this new view on quantum gravity and string theory is that string theory should
be understood as a non-commutative theory in phase space \cite{flmlong1}. From this point of view the usual space-time arena for physics emerges as a maximally degenerate case in an essentially phase space description \cite{flmlong1}. Thus the relation between string theory and non-commutativity is of a fundamental nature. The question of the role of non-associativity in quantum gravity \cite{murat} should be examined from this new viewpoint on string theory.

\medskip

\noindent
\textbf{\large Acknowledgements}
\medskip

This research was supported in part by
 NSERC and MRI ({\small LF}) and
the U.S. Department of Energy ({\small RGL} and {\small  DM}).


\medskip



\end{document}